\documentclass[twocolumn,reprint,aps,floatfix]{revtex4-1}
\usepackage{amsmath,amssymb}
\usepackage{graphicx}
\usepackage{bm}
\usepackage{braket}
\usepackage{physics}
\usepackage{color}
\usepackage{dcolumn}

\begin{document}

\title{Nonequilibrium dynamics in the pump-probe spectroscopy of excitonic insulators}

\author{Tetsuhiro Tanabe$^1$}
\author{Koudai Sugimoto$^2$}
\author{Yukinori Ohta$^1$}
\affiliation{
$^1$Department of Physics, Chiba University, Chiba 263-8522, Japan\\
$^2$Center for Frontier Science, Chiba University, Chiba 263-8522, Japan
}

\date{\today}

\begin{abstract}
We study the nonequilibrium dynamics in the pump-probe spectroscopy of excitonic insulators using 
the spinless two-orbital model with phonon degrees of freedom in the time-dependent mean-field 
approximation.  We introduce the pulse light as a time-dependent vector potential via the Peierls phase 
in the Hamiltonian.  
We find that, in the Bose-Einstein condensation regime where the normal state is semiconducting, the 
excitonic order is suppressed when the frequency of the pulse light is slightly larger than the band gap, 
while the order is enhanced when the frequency of the pulse is much larger than the band gap.  
We moreover find that the excitonic order is completely destroyed in the former situation if the intensity 
of the pulse is sufficiently strong.  In the BCS regime where the normal state is semimetallic, we find 
that the excitonic order is always suppressed, irrespective of the frequency of the pulse light.   
The quasiparticle band structure and optical conductivity spectrum after the pumping are also calculated 
for the instantaneous states.  
\end{abstract}

\maketitle

\section{Introduction}


Nonequilibrium dynamics induced by applying the intense laser pulse have recently been a new way 
of investigating a variety of quantum condensed phases.  Recent achievement of the time resolution 
of a femto-second order enables one to perform experiments studying the ultrafast dynamics of materials.  
Examples include a success of observing light-induced superconductivity \cite{Fausti2011Science, Mitrano2016Nature} 
and a pump-probe measurement of melting of charge-density-wave orders \cite{Perfetti2006PRL, Schmitt2008Science}.

The pump-probe measurement is also applicable to the study of excitonic condensation.  
In the excitonic phase, holes in the valence band and electrons in the conduction band form pairs 
called excitons, just like Cooper pairs of electrons in superconductivity, and they undergo quantum 
condensation at low temperatures \cite{Jerome1967PR, Halperin1968RMP}.  The realization of such 
condensation has been suggested in transition-metal chalcogenides $1T$-TiSe$_2$ 
\cite{Cercellier2007PRL, Kogar2017Science} 
and Ta$_2$NiSe$_5$ \cite{Wakisaka2009PRL, Seki2014PRB}.  Here, we should note that, since the 
spin-singlet excitonic state necessarily couples to the phonon degrees of freedom 
\cite{Phan2013PRB, Zenker2014PRB, Kaneko2013PRB, Sugimoto2016PRB, Kaneko2018PRB}, it is difficult 
to single out the excitonic contributions at least in equilibrium state experiments.  There are, however, 
some attempts to distinguish between the excitonic and phononic contributions using the nonequilibrium 
dynamics induced by laser pulse in 1$T$-TiSe$_2$ 
\cite{Rohwer2011Nature, Mohr-Vorobeva2011PRL, Hellmann2012NC, Monney2016PRB}.  
In Ta$_2$NiSe$_5$, Mor \textit{et al.} found that the band gap can be controlled by the excitation density 
\cite{Mor2017PRL} and argued that its nonequilibrium phenomena come from the exciton dynamics 
\cite{Mor2018PRB}.  Coherent order parameter oscillations caused by the induced phonons were also 
observed \cite{Werdehausen2018JPC, Werdehausen2018SA}.  

The pump-probe spectroscopy experiments in the excitonic phases have been interpreted from the 
theoretical point of view.  While the GW calculations showed that the excitonic order vanishes 
after applying the laser pulse in the BCS regime where the normal state is semimetallic 
\cite{Golez2016PRB}, Murakami \textit{et al.}~\cite{Murakami2017PRL} recently showed that the 
excitonic order can be enhanced by the laser pulse in the Bose-Einstein condensation (BEC) regime 
where the normal state is semiconducting.  Tanaka \textit{et al.}~\cite{Tanaka2018PRB} also showed 
that the switching between the melting and enhancement of excitonic orders can occur when the order 
varies from the BCS regime to BEC regime.  

Note that all of these calculations \cite{Golez2016PRB,Murakami2017PRL,Tanaka2018PRB} assumed 
that the laser pulse excites the electrons in the valence-band orbital directly to the conduction-band 
orbital via the dipole transition.  However, as was discussed in Ref.~[\onlinecite{Wissgott2012PRB}], 
the matrix elements of the dipole transition can be small in the case where the valence-band and 
conduction-band orbitals are well-localized and are spatially separated in distant positions, just 
as in Ta$_2$NiSe$_5$ \cite{Kaneko2013PRB}.  The Peierls term, on the other hand, can survive even 
in such situations \cite{Wissgott2012PRB}.  Thus, there is another way of treating the laser pulse, 
which is to introduce a time-dependent vector potential via the Peierls phase in a tight-binding 
Hamiltonian. 

In this paper, we study the nonequilibrium dynamics of excitonic insulator states applying the 
time-dependent mean-field approximation to the spinless two-orbital model in one-dimension (1D) 
with phonon degrees of freedom, whereby we simulate the situation where an optical laser pulse is 
applied to the system as a pump light. Unlike preceding studies, we here introduce the pulse light 
as a time-dependent vector potential via the Peierls phase in the Hamiltonian of the external field, 
assuming the situations where the dipole matrix elements are small.  We note that the spontaneous 
hybridization between the valence-band and conduction-band orbitals occurs in the symmetry-broken 
excitonic insulator state, so that the interband excitations by the Peierls mechanism can work in 
the present model.  We thus investigate the time evolution of the excitonic order parameter in both 
the BEC and BCS regimes, paying particular attention to its dependence on the frequency and 
intensity of the laser light.  

We will show that, in the BEC regime where the normal state is semiconducting, the excitonic 
order is suppressed when the frequency of the pulse light is slightly larger than the band gap, 
while the order is enhanced when the frequency of the pulse is much larger than the band gap.  
In the BCS regime where the normal state is semimetallic, we will show that the excitonic order 
is always suppressed, irrespective of the frequency of the pulse light.  
We will demonstrate that the excitonic order parameter oscillation occurs in agreement with 
experiment.  We will also calculate the optical conductivity spectrum assuming a single-time, 
instantaneous response for a quasi-steady state after pumping and demonstrate the measurement 
is a useful way for probing the nonequilibrium dynamics of excitonic insulator states.  

The rest of this paper is organized as follows.  
In Sec.~II, we introduce the spinless two-orbital model, define the laser pulse light, and derive 
the equations of motions for the excitonic order parameters in the time-dependent mean-field 
approximation.  In Sec.~III, we present results for nonequilibrium dynamics induced by laser 
pulse in both BEC and BCS regimes.  We also present results for the optical 
conductivity spectra in the nonequilibrium state.  We summarize our results and discuss their 
experimental significance in Sec.~IV.

\section{Model and Method}

\subsection{Spinless two-orbital model}

\begin{figure}[tb]
\begin{center}
\includegraphics[width=0.85\columnwidth]{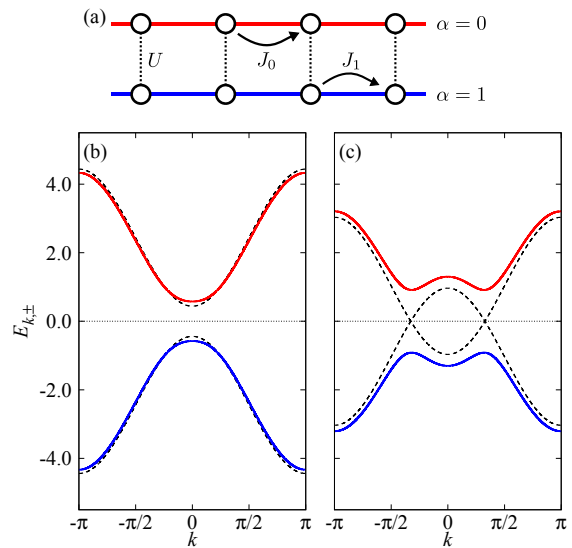}
\end{center}
\caption{
(a) Schematic representation of the spinless two-band model used.  In (b) and (c), we show the 
quasiparticle band dispersions calculated for the model (a) in the equilibrium state (solid lines), 
together with the noninteracting band dispersions (dotted lines).  We assume $D=2.08$ in (b) 
and $D=1.1$ in (c).  
}\label{fig1}
\end{figure}

As a minimum model for describing the spin-singlet excitonic insulator state coupled with 
phonon degrees of freedom, we consider the spinless two-orbital model (or extended 
Falicov-Kimball model \cite{Ihle2008PRB, Zenker2011PRB, Seki2011PRB, Ejima2014PRL}) defined 
on the 1D lattice [see Fig.~\ref{fig1}(a)], interacting with Einstein phonons of frequency 
$\omega_{0}$ \cite{Murakami2017PRL}.  This model may be relevnt to the electronic state of 
an excitonic insulator candidate Ta$_2$NiSe$_5$ with a quasi-1D crystal structure \cite{Seki2014PRB}, 
although the method discussed below is applicable to higher dimensional systems as well.   
Our model is defined by the Hamiltonian 
\begin{equation}
 H = H_\mathrm{e} + H_{\mathrm{e,int}} + H_{\mathrm{ph}} + H_{\mathrm{e-ph}}
 \label{eq:Hamiltonian}
\end{equation}
with
\begin{align}
&H_\mathrm{e}
	= -\sum_{i, \alpha} \big( J_{\alpha}  c^\dagger_{i + 1, \alpha} c_{i, \alpha} + \mathrm{H.c.} \big)
	+ \sum_{i, \alpha} \Delta_{\alpha} c^\dagger_{i, \alpha} c_{i, \alpha}
\label{eq:H_e}
\\
&H_{\mathrm{e,int}} 
	= U \sum_i c^\dagger_{i, 0} c_{i, 0} c^\dagger_{i, 1} c_{i, 1}
\\
&H_{\mathrm{ph}} 
	= \omega_{0} \sum_{i} b^\dagger_{i} b_{i}
\\
&H_{\mathrm{e-ph}} 
	= g \sum_{i} \big( b^\dagger_{i} + b_{i} \big) \big( c^\dagger_{i, 1} c_{i, 0} + \mathrm{H.c.} \big),
\end{align}
where $c_{i, \alpha}$ ($c^\dagger_{i, \alpha}$) is the annihilation (creation) operator of an electron 
at site $i$ with orbital $\alpha$ $(=0,1)$ and $b_{i}$ ($b_i^\dagger$) is the annihilation (creation) operator 
of a phonon at site $i$.  $J_\alpha$, $\Delta_\alpha$, and $U$ are the hopping integral, on-site energy, 
and interorbital repulsive interaction, respectively, and $\omega_0$ and $g$ are the phonon frequency 
and electron-phonon coupling constant, respectively.  Throughout the paper, we set 
$J_\alpha = (-1)^\alpha J$, assuming a direct-gap semiconductor or semimetal [see Figs.~\ref{fig1}(b) 
and \ref{fig1}(c)], so that we create a momentum $q=0$ condensate appropriate to Ta$_2$NiSe$_5$, 
rather than $q=\pi$ assumed in, e.g., Ref.~[\onlinecite{Golez2016PRB}].  
We keep the relation $U=-(\Delta_0+\Delta_1)$, so that the condition that the number of electrons 
per site (containing two orbitals) is one is satisfied at the chemical potential $\mu=0$.  
We define the energy level difference $D=\Delta_0-\Delta_1$.  We introduce the effective 
electron-phonon interaction $\lambda=2g^2/\omega_0$ and assume the values $\lambda=0.1$ and 
$\omega_0=0.1$ throughout the paper.  We use the values $J=1$ (unit of energy), $\Delta_0=-0.36$, 
$\Delta_1=-2.44$, $D=2.08$, and $U=2.8$ unless otherwise indicated.  The noninteracting band dispersions 
used are illustrated in Figs.~\ref{fig1}(b) and \ref{fig1}(c).  

\subsection{Optical laser pulse}

To treat the optical laser pulse as a pump light, we introduce the time-dependent vector potential as a 
Peierls phase to the Hamiltonian.  We assume that the vector potential is approximately independent of 
the spatial position since the wavelength of the light is much longer than the lattice spacing $a$.  
The hopping-integral term in Eq.~(\ref{eq:H_e}) is then replaced by \cite{Freericks2017PS}
\begin{equation}
 J_{\alpha}  c^\dagger_{i + 1, \alpha} c_{i, \alpha}
 \to
 J_{\alpha} e^{i ea A(t)}  c^\dagger_{i + 1, \alpha} c_{i, \alpha},
\end{equation}
where $e$ $(<0)$ is the elementary charge, for which we set $e=-1$.  
We use the units $\hbar = c = 1$ 
and the lattice constant $a=1$ throughout the paper.  
We consider the Gaussian-type laser pulse defined by the time-dependent vector potential as 
\cite{Matsueda2012JPSJ} 
\begin{equation}
 A (t)
	= \theta (t) A_0 e^{- \frac{\left( t - t_{\mathrm{p}} \right)^2}{2 \sigma_{\mathrm{p}}^2}} \sin \Omega t  ,
\end{equation}
where $\Omega$ and $A_0$ are the frequency and intensity of the pulse light, respectively, and 
\begin{equation}
\theta (t) = 
 \begin{cases}
	0 & (t \leq 0) \\
	1 & (t > 0)
\end{cases}
\end{equation}
is the step function.  
The time and width of the pulse are described by $t_{\mathrm{p}}$ and $\sigma_{\mathrm{p}}$, respectively.  
In this paper, we set $\sigma_{\mathrm{p}} = 30$ and $t_{\mathrm{p}} = 100$.  
By the Fourier transformation $c_{i, \alpha} = \frac{1}{\sqrt{N}} \sum_{k} e^{i k r_i} c_{k, \alpha}$, where 
$N$ is the number of lattice sites in the system, Eq.~(\ref{eq:H_e}) may be rewritten as 
\begin{equation}
 H_\mathrm{e}
	= \sum_{k, \alpha} \varepsilon_{\alpha} ( ka - ea A(t) )  c^\dagger_{k, \alpha} c_{k, \alpha}
\end{equation}
where $\varepsilon_{\alpha} (ka) = -2 \left( -1 \right)^\alpha \cos ka + \Delta_{\alpha}$.  

\subsection{Equations of motion}

The time evolution of the system induced by the laser pulse is calculated in the time-dependent 
mean-field approximation \cite{Murakami2017PRL, Barankov2004PRL, Yezbashyan2005PRB}.  
We define the uniform excitonic order parameter as 
$\phi (t) = \langle c^\dagger_{i, 0} (t) c_{i, 1} (t) \rangle$, 
the uniform phonon displacement as 
$X (t) = \langle b^\dagger_i (t) + b_i (t) \rangle$, and 
the uniform electron density as 
$n_\alpha (t) = \langle c^\dagger_{i, \alpha} (t) c_{i, \alpha} (t) \rangle$.  
The time dependence of the operators is given by the Heisenberg representation.  

For convenience, we use the pseudospin representation for the spinless electron in the two orbitals 
\cite{Murakami2017PRL}.  We define the pseudospin as 
$
S_k^\gamma = \frac{1}{2} \bm{\Psi}_k^\dagger \sigma_\gamma \bm{\Psi}_k
$,
where 
$
\bm{\Psi}_k^\dagger
=\begin{pmatrix}
 c^\dagger_{k, 0} & c^\dagger_{k, 1}
\end{pmatrix}
$
is the spinor, $\sigma_\gamma$ ($\gamma = x,y,z$) is the $\gamma$ component of Pauli matrix 
$\bm{\sigma}$, and $\sigma_0$ is the identity matrix.
Introducing the pseudomagnetic field as 
\begin{subequations}
\begin{align}
&B^x_k (t)
	= 2g X (t) - 2 U \mathrm{Re} \, \phi (t)
\\
&B^y_k (t)
	= - 2 U \mathrm{Im} \, \phi (t)
\\
&B^z_k (t)
	=  \left[ \varepsilon_{0} ( ka - ea A(t) )  - \varepsilon_{1} ( ka - ea A(t) )  \right] \notag \\
		&~~~~~~~~~~- U \left( n_0 (t) - n_1 (t) \right)
\\
&B^0_k (t)
	= \sum_\alpha \left[ \varepsilon_\alpha (ka-eaA(t)) + U n_{\alpha} (t) \right],
\end{align}
\label{eq:pseudoB}
\end{subequations}
we may write the mean-field Hamiltonian as 
\begin{equation}
 H^{\mathrm{MF}} (t) =  H^{\mathrm{MF}}_\mathrm{e} (t) + H^{\mathrm{MF}}_{\mathrm{ph}} (t)
\end{equation}
with
\begin{align}
&H^{\mathrm{MF}}_\mathrm{e} (t)
	= \sum_{k} \sum_{\gamma=0,x,y,z} B^{\gamma}_k (t) S^{\gamma}_k
\label{eq:H^MF_e}
 \\
&H^{\mathrm{MF}}_{\mathrm{ph}} (t) =
 \omega_0 \sum_{i} b_i^\dagger b_i + 2g \, \mathrm{Re} \, \phi (t) \sum_{i} \big( b_i + b_i^\dagger \big).
\end{align}

 
The time-dependent variables $\phi (t)$, $X (t)$, and $n_{\alpha} (t)$ may be calculated 
from the Heisenberg equations of motion.  We thus obtain the equations 
\begin{subequations}\begin{align}
&\frac{\partial \left\langle \bm{S}_k (t) \right\rangle }{\partial t}
	= \bm{B}_k (t) \times \left\langle \bm{S}_k (t) \right\rangle
\\
&\frac{\partial \left\langle S^0_k (t) \right\rangle }{\partial t}
	= 0
\\
&\frac{\partial X (t) }{\partial t}
	= \omega_0 P (t)
\label{eq:EOM_X}
\\
&\frac{\partial P (t) }{\partial t}
	= - \omega_0 X (t) - 4g \, \mathrm{Re} \, \phi (t) , 
\label{eq:EOM_P}
\end{align}\label{eq:EOM}\end{subequations}
where $P (t) = i \langle b^\dagger_i - b_i \rangle$ is the momentum of the phonon.  
We solve Eqs.~(\ref{eq:EOM}) numerically using the Runge-Kutta fourth-order method, and 
substitute the solutions into 
\begin{equation}
 \begin{pmatrix}
	n_0 (t) & \phi^* (t) \\
	\phi (t) & n_1 (t)
 \end{pmatrix}
	= \frac{1}{N}  \sum_k \left[  \left\langle \bm{S}_k (t) \right\rangle \cdot \bm{\sigma} + \left\langle S^0_k (t) \right\rangle \sigma_0  \right]
\label{eq:order_parameters_and_pseudospins}
\end{equation}
to obtain the excitonic order parameter and the number of electrons.  

\subsection{Equilibrium state}

To solve the above differential equations, we need to set the initial conditions.  
At $t = 0$, where the external field is absent, the system is in equilibrium.  
Since we consider the system at zero temperature, the phonons are at rest, or $P(0) = 0$.
From Eq.~(\ref{eq:EOM_P}), the phonons satisfy $X(0) = - \frac{4g}{\omega_0} \phi (0)$.  
We note that the excitonic order parameter is real when the electron-phonon coupling 
is present \cite{Zenker2014PRB,Kaneko2015PRB}.  

Diagonalizing Eq.~(\ref{eq:H^MF_e}), we obtain
\begin{equation}
 H^{\mathrm{MF}}_\mathrm{e} (t)
	= \sum_k \big( 
	E_{k, +} (t) \gamma^\dagger_{k, +} \gamma_{k, +}
	+ E_{k, -} (t) \gamma^\dagger_{k, -} \gamma_{k, -}
	\big),
\end{equation}
where $\gamma_{k, \pm}$ ($\gamma^\dagger_{k, \pm}$) is the annihilation (creation) 
operator of the quasiparticle with dispersions 
$E_{k, \pm} (t) = \left[ B^0_k (t) \pm \left| \bm{B}_k (t) \right| \right]/2$.  
At $t=0$, the expectation value of the pseudospin is given by 
\begin{align}
& \left\langle S^\gamma_k (0) \right\rangle = \nonumber \\
& \begin{cases}
	\displaystyle{\frac{B^\gamma_k(0)}{2 \left| \bm{B}_k (0) \right|}} \left[ f (E_{k, +}(0)) - f (E_{k, -}(0))  \right]
	& (\gamma = x,y,z)
\\
 	\displaystyle{\frac{1}{2}} \left[f (E_{k, +} (0)) + f (E_{k, -} (0)) \right]
 	 & (\gamma = 0)
\end{cases}
\label{eq:self-consistent_equation}
\end{align}
where $f (E)$ is the Fermi distribution function.  
Solving the self-consistent equations for the pseudospins, we obtain Eq.~(\ref{eq:self-consistent_equation}) 
with Eqs.~(\ref{eq:pseudoB}) and (\ref{eq:order_parameters_and_pseudospins}) at $t=0$, which we use as 
the initial conditions for Eqs.~(\ref{eq:EOM}).  
The calculated quasiparticle band dispersions at $t=0$ are shown in Figs.~\ref{fig1}(b) and \ref{fig1}(c).  


\section{Results of calculations}

Here, we discuss the results of calculations mainly when the system is semiconducting 
in the normal phase.  Our parameter values give the quasiparticle band gap of size 1.15 
in the presence of the excitonic order, which is considerably enhanced in comparison to 
the band gap of a size 0.88 in the absence of the excitonic order [see Fig.~\ref{fig1}(b)].  
We also discuss the results when the system is semimetallic in the last part of this section.  
All the calculations are made at absolute zero temperature.  

\begin{figure}[tb]
\begin{center}
\includegraphics[width=1.0\columnwidth]{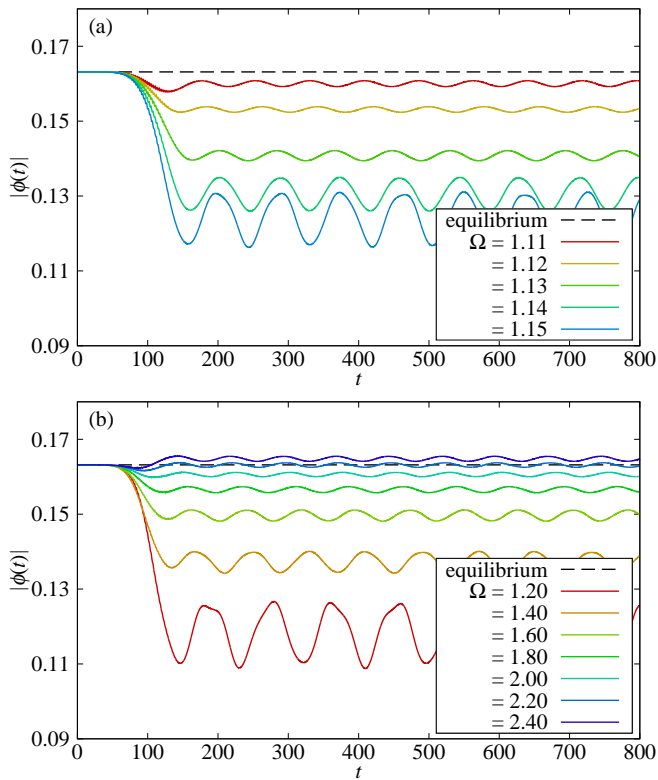}
\end{center}
\caption{
Calculated time evolution of the absolute value of the excitonic order parameter $|\phi(t)|$ 
for a variety of frequency of the laser pulse $\Omega$ at $A_0=0.05$.  The BEC (or semiconducting) 
regime is assumed.  (a) The case where $\Omega$ is smaller than the quasiparticle band gap $1.15$ 
and (b) the case where $\Omega$ is larger than the quasiparticle band gap.
}\label{fig2}
\end{figure}

\begin{figure}[tb]
\begin{center}
\includegraphics[width=1.0\columnwidth]{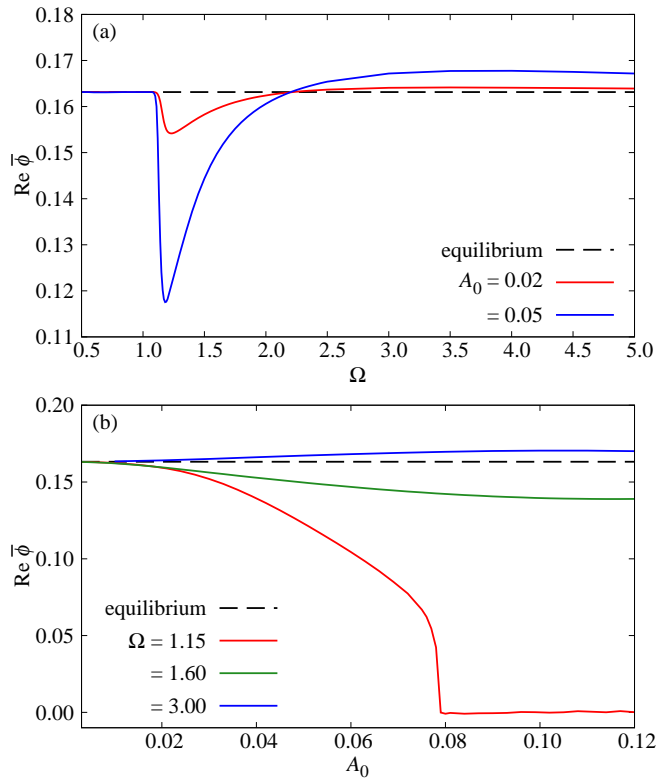}
\end{center}
\caption{
Time averages of the real part of the excitonic order parameter $\mathrm{Re}\,\bar{\phi}$ 
calculated as a function of (a) the frequency of the laser light $\Omega$ and (b) intensity of 
the laser light $A_0$.  
}\label{fig3}
\end{figure}

\begin{figure}[tb]
\begin{center}
\includegraphics[width=1.0\columnwidth]{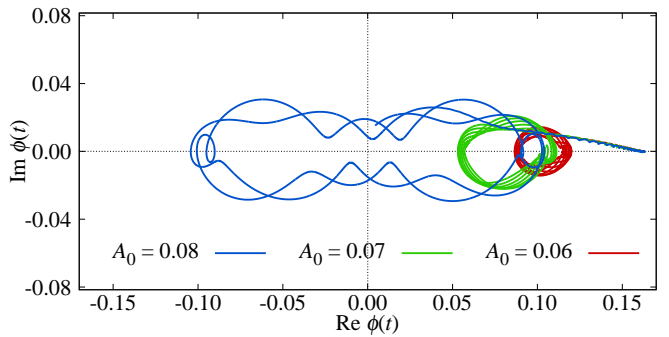}
\end{center}
\caption{
Calculated trajectory of the excitonic order parameter in the complex plane for $0\le t<800$.  
The laser pulse is applied at $t=0$, where $\mathrm{Re}\,\phi(t)=0.163$ and 
$\mathrm{Im}\,\phi(t)=0$.  We assume $\Omega=1.15$.  
}\label{fig4}
\end{figure}

\subsection{Evolution of the order parameter}

First, let us discuss the evolution of the excitonic order parameter.  
In Fig.~\ref{fig2}, we show the results for the time evolution of the excitonic order parameter.  
We find in Fig.~\ref{fig2}(a) that, when the frequency of the laser light is smaller than the band 
gap, the order parameter remains unchanged against the laser pulse.  This is because the 
pulse cannot excite the electrons in the valence band to the conduction band.  
Once the frequency of the laser light becomes comparable to the size of the band gap, the 
excitations of electrons occur and the order parameter drastically changes as seen in 
Figs.~\ref{fig2}(a) and \ref{fig2}(b).  After the laser pulse passes through the system, the 
absolute value of the order parameter decreases rapidly in comparison to the equilibrium 
value, and it oscillates with the frequency of the phonon.  This result indicates that the 
excitonic order partially melts by the laser pulse.  
Note that, since the electron-electron correlations and electron-phonon couplings are treated 
in the mean-field approximation, the thermalization process of the nonequilibrium state after 
the pulse light passes through the system 
\cite{Tsuji2013PRL,Kemper2015PRB,Sentef2016PRB,Schuler2018PRL} is excluded 
in our calculations.  As a result, the order parameter oscillates persistently without damping.  

On the other hand, when the frequency of the laser light is much larger than the size of the 
band gap, we find that the opposite result occurs.  As seen in Fig.~\ref{fig2}(b), we find that, 
after the laser pulse passes through the system, the absolute value of the order parameter 
is enhanced.  
This result can be explained as the effect of the Hartree shift \cite{Murakami2017PRL}: 
The strong excitations induced by the pulse lead to the reduction of the number of valence 
electrons and increase in the number of conduction electrons, which reduces the size of the 
band gap by the Hartree shift, and therefore the excitonic order parameter is enhanced.  

To see the dependence of the excitonic order parameter on the frequency and intensity of 
the laser pulse more clearly, we introduce the time average of the order parameter 
$\bar{\phi} = \frac{1}{t_2 - t_1} \int^{t_2}_{t_1} \phi(t) dt$, 
which is calculated in a sufficiently long time interval between $t_1$ and $t_2$ 
after the pulse is applied.  The oscillations due to the phonons are thus obliterated.  
In Fig.~\ref{fig3}(a), we show the real part of the averaged order parameter $\textrm{Re}\,\bar{\phi}$ 
as a function of the frequency of the laser light.  We find that, while the order parameter 
is suppressed for the frequencies slightly larger than the band gap, it is enhanced for the 
frequencies much larger than the band gap.  This result can be interpreted as the competition 
between the melting of the excitonic order and the reduction of the band gap caused by the 
Hartree shift.

We also calculate the real part of the averaged order parameter as a function of the intensity 
of the laser light.  As shown in Fig.~\ref{fig3}(b), we find that, when the frequency of the light 
is around the size of the band gap, there appears a critical value of the intensity at which the 
averaged order parameter completely vanishes, retaining only the oscillation of the phonons.  
Note that, when the frequency of the laser light is much larger than the band gap, such 
suppression of the excitonic order does not occur.  

Although not shown here, we also calculate the time average of the phonon displacement 
$\bar{X}$ and find that the behavior is very similar to that of the order parameter $\bar{\phi}$ 
shown in Fig.~\ref{fig3}.  

We can understand the present result clearly from the trajectory of the excitonic order 
parameter in the complex plane.  The results are shown in Fig.~\ref{fig4}, where we find 
the following:  At $t=0$, the excitonic order parameter is a real number since the electron-phonon 
coupling term in the Hamiltonian fixes the phase of the order parameter to zero 
\cite{Zenker2014PRB, Kaneko2015PRB}.  After the laser pulse is applied, the excitonic order 
parameter goes around a certain point where the imaginary part is zero and the real part is smaller 
than that in the equilibrium state.  As the intensity of the laser pulse increases, the central point 
around which the order parameter oscillates approaches the origin, and finally it reaches 
the origin when the intensity become larger than the critical value.  Thus, the strong laser pulse 
destroys the excitonic order.  

\begin{figure}[tb]
\begin{center}
\includegraphics[width=0.95\columnwidth]{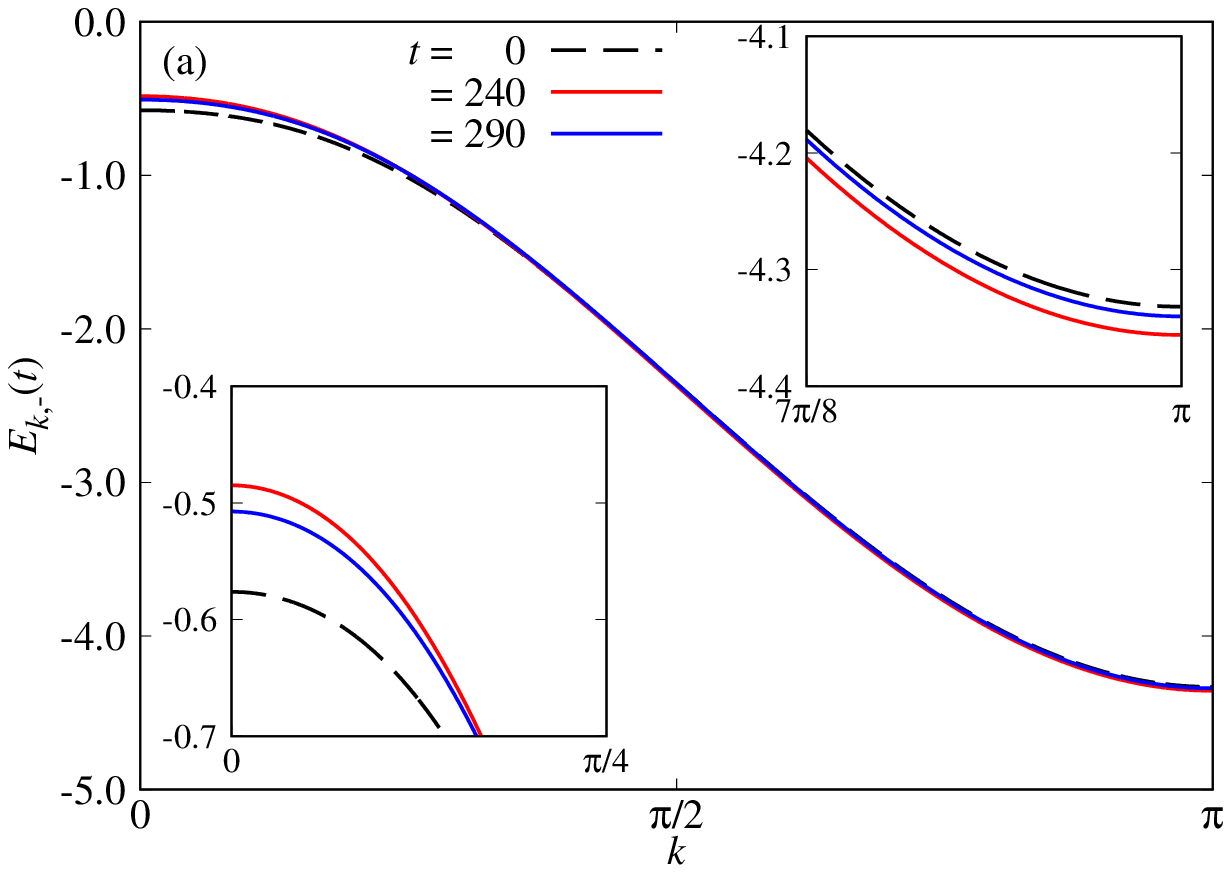}
\includegraphics[width=0.95\columnwidth]{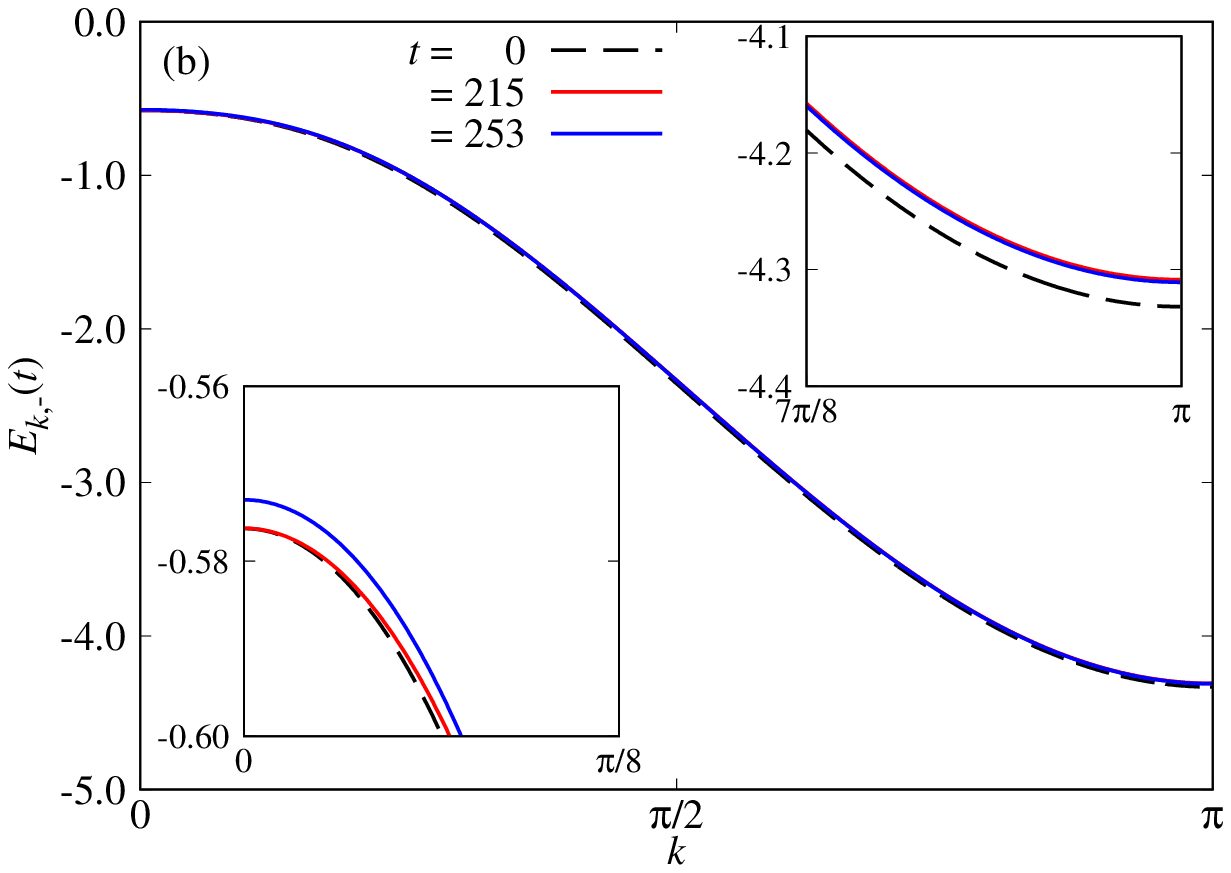}
\end{center}
\caption{
Quasiparticle band dispersion below the Fermi level $E_{k,-}(t)$ calculated at time $t$.  
We assume (a) $A_0=0.05$ and $\Omega=1.15$ and (b) $A_0=0.05$ and $\Omega=3$.  
The dashed line indicates the quasiparticle band dispersion at $t=0$ (or in the equilibrium 
state).  Insets enlarge the dispersions.  At $t>0$, the dispersion oscillates between 
the red solid line and blue solid line.  
}\label{fig5}
\end{figure}

\subsection{Quasiparticle band dispersion}

We calculate the quasiparticle band dispersion in the time-dependent mean-field approximation 
of the model after the laser pulse is applied.  The time- and angle-resolved photoemission 
spectroscopy experiment can in principle observe this dispersion.  The results are illustrated in 
Figs.~\ref{fig5}(a) and \ref{fig5}(b).  We find in Fig.~\ref{fig5}(a) that the size of the energy gap 
decreases when the frequency of the laser light is slightly larger than the energy gap 
(or at $\Omega=1.15$).  This is because the laser pulse leads to the partial melting of the 
excitonic order as shown in Fig.~\ref{fig3}(b), of which the behavior seems consistent with recent 
experiment \cite{Mor2017PRL}.  

We then find in Fig.~\ref{fig5} (b) that the size of the energy gap also decreases even when the 
frequency of the laser light is much larger than the energy gap (or at $\Omega=3$).  
Thus, the top of the valence band goes up in the entire frequency region of the pulse light.  
However, the bottom of the valence band behaves differently depending on the frequency of the 
laser light: it goes down at $\Omega=1.15$ but it goes up at $\Omega=3$.  

\begin{figure}[tb]
\begin{center}
\includegraphics[width=0.95\columnwidth]{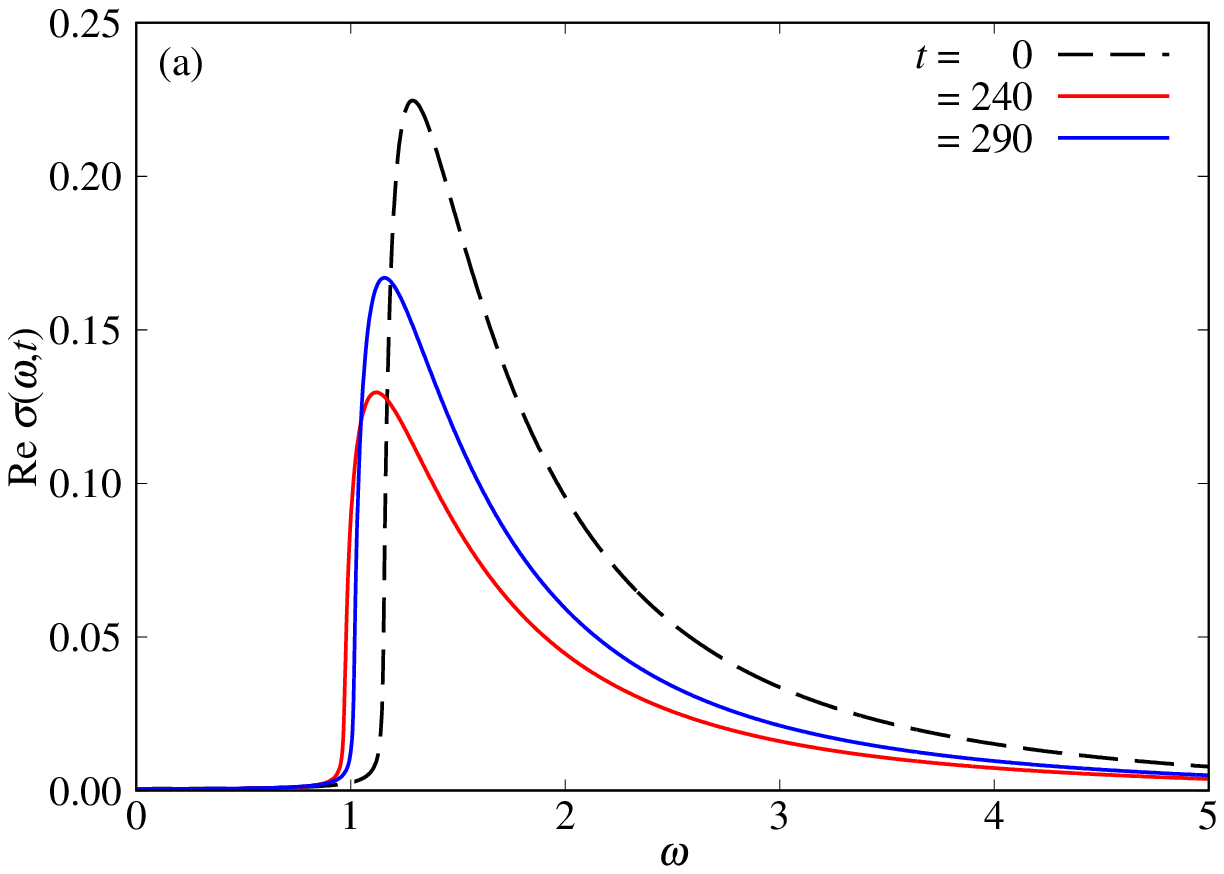}
\includegraphics[width=0.95\columnwidth]{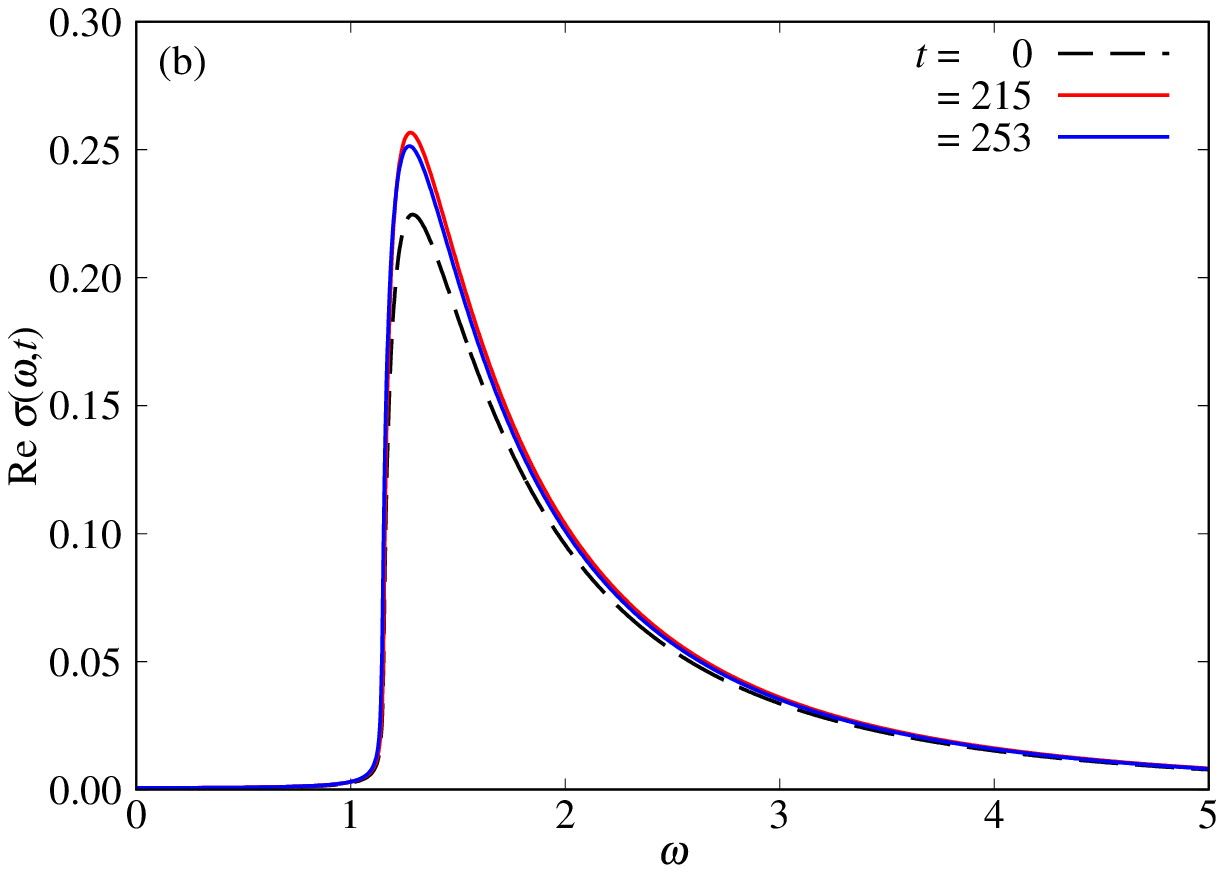}
\end{center}
\caption{
Real part of the optical conductivity spectrum $\mathrm{Re}\,\sigma(\omega,t)$ calculated 
at time $t$.  The dashed line indicates the optical conductivity spectrum at $t=0$ (or in the 
equilibrium state).  We assume (a) $A_0=0.05$ and $\Omega=1.15$ and (b) $A_0=0.05$ and 
$\Omega=3$.  At $t>0$, the spectrum oscillates between the red solid line and blue solid line.  
}\label{fig:optcond}
\end{figure}

\subsection{Optical conductivity spectrum}

To make a connection between our theory and experiment, we calculate the real part of the 
optical conductivity spectrum $\mathrm{Re} \, \sigma (\omega,t)$ measured at the frequency 
$\omega$ of the probe light after the pulse light passes through the system.  The assumption 
used here is that the double-time retarded response that is actually required in a true 
pump-probe setting \cite{Eckstein2008PRB, Lenarcic2014PRB, Shao2016PRB} can be 
approximated by a single-time, instantaneous response for a quasi-steady state after 
pumping \cite{Fukaya2015NC}.  In this approximation, the real part of the optical conductivity 
of the present nonequilibrium state at a certain moment $t$ is given by
\begin{multline}
 \mathrm{Re} \, \sigma (\omega,t)
	= -\frac{\pi}{N \omega}
		\sum_{k}
		\left[ f (E_{k, +} (t)) - f (E_{k, -} (t)) \right]
\\ \times
		\left| J_{+-} (k,t) \right|^2
		\delta (E_{k, +}(t) - E_{k, -}(t) - \omega) , 
\end{multline}
where 
\begin{equation}
J_{+-}(k,t) = 2ea \frac{B^x_k(t)+iB^y_k(t)}{\left| \bm{B}_k (t) \right|} \sin ka
\end{equation}
is the off-diagonal matrix element of the electric current.  
When the excitonic order is absent, the optical conductivity completely vanishes in the 
present model.  This is because the orbital off-diagonal elements of the current operator 
are all zero due to the orthogonality of the orbitals, prohibiting the excitations of electrons 
from the valence to the conduction bands.  Only when the excitonic order appears, or the 
hybridization between the valence and conduction bands occurs, the optical conductivity 
acquires the finite spectral weight.  

The calculated results are shown in Fig.~\ref{fig:optcond} at $\Omega=1.15$ and $\Omega=3$.  
We find that the optical conductivity spectrum acquires the finite spectral weight for the 
probe frequency $\omega$ larger than the quasiparticle band gap.  
At $\Omega=1.15$, where the excitonic order is suppressed and the size of the band gap 
is reduced, we find that the peak position of the optical conductivity spectrum shifts to 
lower energy side and the peak height decreases in comparison to that of the equilibrium 
state [see Fig.~\ref{fig:optcond}(a)].  
At $\Omega=3$, where the excitonic order is enhanced but the size of the band gap remains 
nearly unchanged, we find that the peak position of the optical conductivity spectrum remains 
unchanged but the peak height slightly increases in comparison to that of the equilibrium 
state [see Fig.~\ref{fig:optcond}(b)].  

We note again that, in the actual measurement of the optical conductivity in the nonequilibrium 
state, the spectra may be affected by the phonon oscillations, so that the double-time retarded 
response should be taken into account in future improved calculations.  

\begin{figure}[thb]
\begin{center}
\includegraphics[width=1.0\columnwidth]{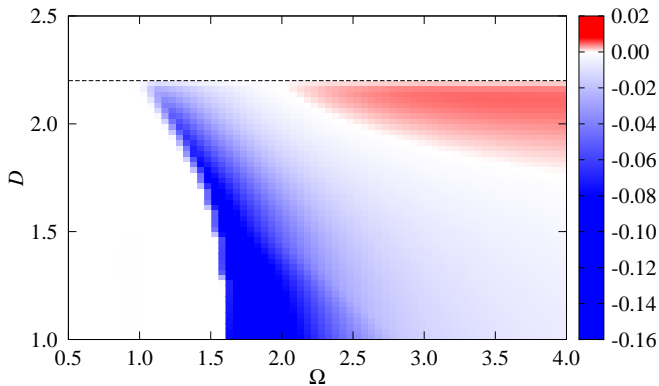}
\end{center}
\caption{
Contour plot of the time average of the real part of the excitonic order parameter 
${\rm Re}\,\bar{\phi}$ calculated in the parameter space of $(D,\Omega)$ at $U=2.8$.  
We assume $A_0=0.05$.  The excitonic order is suppressed (enhanced) in the blue 
(red) region of the parameter space.  
}\label{fig7}
\end{figure}

\subsection{Semimetallic case}

Finally, let us discuss the case where the quasiparticle band structure is semimetallic in the 
normal state (or in the absence of the excitonic order).   To see this, we calculate the time 
average of the real part of the excitonic order parameter ${\rm Re}\,\bar{\phi}$ as a function 
of the level difference $D=\Delta_0-\Delta_1$.  The semiconducting and semimetallic regions 
in the normal state are separated at $D=1.61$.  The calculated results are summarized as a 
phase diagram, which is shown in Fig.~\ref{fig7}.  We find that the excitonic order is not stable 
at $D>2.20$, that either the suppression or enhancement of the excitonic order occurs 
depending on the frequency of the laser light in the semiconducting region (or at $1.61<D<2.20$) 
as shown in Fig.~\ref{fig3}(a), and that only the suppression of the order parameter occurs 
in the semimetallic region (or at $D<1.61$).  This result is consistent with the preceding 
study assuming the semimetallic band structure \cite{Golez2016PRB}.  

\section{Summary and Discussion}

In summary, we have studied the nonequilibrium dynamics of excitonic insulator states using the 
spinless two-orbital model with phonon degrees of freedom in the time-dependent mean-field 
approximation.  Unlike preceding studies where the dipole transition of electrons between 
the valence and conduction bands was assumed, we have introduced the pulse light as a 
time-dependent vector potential via the Peierls phase in the Hamiltonian.  

We have then found that, in the BEC regime where the normal state is semiconducting, the 
excitonic order is suppressed when the frequency of the pulse light is slightly larger than the 
band gap, while the order is enhanced when the frequency of the pulse is much larger than 
the band gap.  
We have moreover found that the excitonic order is completely destroyed in the former situation 
if the intensity of the pulse is sufficiently strong.  In the BCS regime where the normal state 
is semimetallic, we have found that the excitonic order is always suppressed, irrespective 
of the frequency of the pulse light.   We have also calculated the time-dependent quasiparticle 
band dispersion and optical conductivity spectrum of the model.  

Finally, let us discuss possible experimental significance of our results, taking a quasi-1D 
direct-gap semiconductor Ta$_2$NiSe$_5$ as an example.  The energy of the laser light so far 
used in experiment is 1.55 eV 
\cite{Mor2017PRL,Mor2018PRB,Werdehausen2018JPC,Werdehausen2018SA}, 
which is very large in comparison with the observed band gap 160 meV 
\cite{Seki2014PRB,Lu2017NC,Larkin2017PRB}.  Thus, our two-band model seems to be too 
simple to take into account the excitations in such a high-energy scale where other orbitals 
such as Se $4p$ become relevant.  In other words, the use of lower-energy laser lights may 
be more informative to clarify the nonequilibrium dynamics of Ta$_2$NiSe$_5$ in the lowest 
energy scales.  
Also, it was pointed out that Ta$_2$NiSe$_5$ is in the strong coupling BEC regime, despite 
the fact that the noninteracting band structure is semimetallic \cite{Sugimoto2018PRL}.  
Such a situation cannot be treated in our simple mean-field approximation.  
Nevertheless, we may point out that our calculations could yield some experimental 
aspects such as the enhancement of the excitonic order \cite{Mor2017PRL,Mor2018PRB}, 
coherent order parameter oscillations \cite{Werdehausen2018JPC,Werdehausen2018SA}, 
as well as the insulator-to-metal transitions reported recently \cite{Okazaki2018NC}.  
The improved method of calculations beyond the mean-field approximation based on more 
realistic models will be required for future quantitative studies of the nonequilibrium dynamics 
of the excitonic insulator states.  
We may note here that, in the case of superconductivity under periodically oscillating phonons 
driven by an external field, the order parameter is reduced in the dynamical mean-field-theory 
calculation \cite{Murakami2017PRB}, while it is enhanced in the effective electron-electron 
interaction scheme \cite{Knap2016PRB, Komnik2016EPJB}; the opposite results are obtained 
depending on the approximations used.  Thus, a careful treatment of strongly interacting 
systems will be required in future calculations of nonequilibrium phenomena observed in the 
excitonic phases as well.  

\section*{Acknowledgments}
We thank T. Kaneko, T. Mizokawa, Y. Murakami, K. Okazaki, Y. Yamada, and K. Yonemitsu for 
enlightening discussions.  This work was supported in part by a Grant-in-Aid for Scientific 
Research (No.~JP17K05530) from JSPS of Japan.

\end{document}